\documentstyle[12pt]{article}

\newcommand{\be}{\begin{equation}}
\newcommand{\ee}{\end{equation}}
\newcommand{\ba}{\begin{eqnarray}}
\newcommand{\ea}{\end{eqnarray}}
\newcommand{\bas}{\begin{eqnarray*}}
\newcommand{\eas}{\end{eqnarray*}}

\newcommand{\nn}{\nonumber}

\newcommand{\g}{\gamma}
\newcommand{\Lam}{\Lambda}
\newcommand{\th}{\theta}
\newcommand{\dl}{\delta}

\newcommand{\cA}{{\cal A}}
\newcommand{\cB}{{\cal B}}
\newcommand{\cC}{{\cal C}}
\newcommand{\cD}{{\cal D}}

\begin{document}
 
\addtolength{\baselineskip}{0.20\baselineskip}
\hfill UT-Komaba 96-31
 
\hfill December 1996
\begin{center}
 
\vspace{36pt}
{\large \bf  Boundary $S$ matrices for \\
the open Hubbard chain with boundary  fields }
 
\end{center}

\vspace{36pt}
 
\begin{center}

\vspace{5pt}
 Osamu Tsuchiya \footnote{E-mail address: otutiya@hep1.c.u-tokyo.ac.jp} \\
\vspace{3pt}
{\sl Department of Pure and Applied Sciences, 
University of Tokyo, \\
Komaba, Meguro-ku, Tokyo 153, Japan}

\end{center}

\vspace{36pt}
 
 
\begin{center}
{\bf Abstract}
\end{center}
 
\vspace{12pt}
 
\noindent
Using the method introduced by Grisaru et al.,
boundary $S$ matrices for the physical excitations
of the open Hubbard chain with boundary  fields
are studied.
In contrast to the open supersymmetric $t$-$J$ model,
the boundary $S$ matrix for the charge excitations
depend on the boundary fields
though the boundary fields 
do not break the spin-$SU(2)$ symmetry.
 
\vspace{24pt}
 
\vfill
 
\newpage
 

Recently, one-dimensional integrable models {\it with boundaries}
have attracted renewed interest.
Those models provide relevant informations for the boundary effects
on the one-dimensional strongly correlated systems.
Among others, as for the bulk case, 
the one-dimensional Hubbard model with open boundary conditions
(open Hubbard chain) plays an important role in this field.

In this letter, 
using the method introduced 
by Grisaru et al. \cite{GMN} (see also ref. \cite{Ess96}),
we study the boundary $S$ matrix for quasiparticles
of the open Hubbard chain with boundary  fields.  

Let us first recall the known facts about the (open) Hubbard chain.
The Hamiltonian of the open Hubbard chain 
with boundary fields is given by 
$H^{(\pm)}
=H_{\mbox{{\scriptsize bulk}}}^{(\mbox{{\scriptsize open}})}
+H_{\mbox{{\scriptsize boundary}}}^{(\pm)}$,
where 
\ba
\label{bulk-hamil}
H_{\mbox{{\scriptsize bulk}}}^{(\mbox{{\scriptsize open}})} 
&=& 
-\sum_{j=1}^{L-1}\sum_{\sigma=\uparrow,\downarrow}
(\psi_{j \sigma}^{\dagger}\psi_{j+1 \sigma} 
+\psi_{j+1 \sigma}^{\dagger}\psi_{j \sigma})
+ 
U\sum_{j=1}^{L}
(n_{j \uparrow}-1/2)(n_{j\downarrow}-1/2), \\ 
\label{boundary-hamil}
H_{\mbox{{\scriptsize boundary}}}^{(\pm)}
&=&
-h_{1}(n_{1\uparrow}\pm n_{1\downarrow})
-h_{L}(n_{L\uparrow}\pm n_{L\downarrow}). 
\ea
Here $U$ is  the coupling constant, 
$h_l$ is the boundary field at site $l\in\{1, L\}$,
$\psi_{j \sigma}$ (resp. $\psi_{j \sigma}^{\dagger}$)
denotes the annihilation (resp. the creation) operator
of an electron with spin $\sigma\in\{\uparrow, \downarrow\}$
at site $j\in\{1, 2, \cdots, L\}$,
and $n_{j \sigma}=\psi_{j \sigma}^{\dagger}\psi_{j \sigma}$
is the number operator.

It is well known that the bulk Hamiltonian 
$H_{\mbox{{\scriptsize bulk}}}^{(\mbox{{\scriptsize open}})}$ 
(on the bipartite lattice, {\it i.e.}, with even $L$)
possesses an $SO(4)\cong (SU(2)\times SU(2))/\mbox{{\bf Z}}_2$ 
symmetry \cite{eta} (see also ref. \cite{EKS92}).
That is, together with the ordinary spin-$SU(2)$ symmetry
which corresponds to the spin degrees of freedom, 
$H_{\mbox{{\scriptsize bulk}}}^{(\mbox{{\scriptsize open}})}$ 
is also invariant under the action of 
the so-called $\eta$-$SU(2)$ algebra
which pertains to the charge degrees of freedom.
The boundary Hamiltonian $H_{\mbox{{\scriptsize boundary}}}^{(+)}$
(resp. $H_{\mbox{{\scriptsize boundary}}}^{(-)}$) breaks
the $\eta$-$SU(2)$ symmetry (resp. the spin-$SU(2)$ symmetry)
down to $U(1)$. 
Quasiperticle spectra of the attractive Hubbard model 
and those of the repulsive Hubbard model
are related by an interchange of 
the spin and charge degrees of freedom \cite{EK94}.
Then, in what follows, we restrict attention to the Hamiltonian
$H^{(+)}$ with $U>0$ (repulsive case).

The open Hubbard chain with boundary fields 
has been solved by 
the (coordinate) Bethe ansatz method \cite{S85,AS,DY}.
The Bethe ansatz for this model provides eigenstates of 
the Hamiltonian $H^{(+)}$ 
which are parameterized by the two sets of roots ('rapidities')
$\{k_j\}_{j=1}^N$ and $\{\Lam_\gamma\}_{\gamma=1}^M$.
Here $N$ is the number of electrons
and $M$ is the number of electrons with down spin.
These roots are subject to the (nested) Bethe ansatz equations,
\ba
\label{bethe-eq1} 
e^{i2k_{j}(L+1)} 
\beta(k_{j},h_{1}) \beta(k_{j},h_{L})=  
\prod_{\delta =1}^{M}
\frac{(\Lam_{\dl}-\sin k_{j}-ic/2)(\Lam_{\dl}+\sin k_{j}+ic/2)}
     {(\Lam_{\dl}-\sin k_{j}+ic/2)(\Lam_{\dl}+\sin k_{j}-ic/2)},&&\\
\label{bethe-eq2} 
\prod_{{\scriptstyle \dl=1} \atop {\scriptstyle \dl (\neq \g)}}^{M}
\frac{(\Lam_{\g}-\Lam_{\dl}-ic)(\Lam_{\g}+\Lam_{\dl}-ic)}
     {(\Lam_{\g}-\Lam_{\dl}+ic)(\Lam_{\g}+\Lam_{\dl}+ic)}
=
\prod_{j=1}^{N} 
\frac{(\Lam_{\g}-\sin k_{j}-ic/2)(\Lam_{\g}+\sin k_{j}-ic/2)}
     {(\Lam_{\g}-\sin k_{j}+ic/2)(\Lam_{\g}+\sin k_{j}+ic/2)},&& 
\ea 
where $j=1, \cdots, N,$ $\g=1, \cdots, M,$ and 
\ba
&&
c=U/2,  \\
&&
\beta(x,h)=\frac{1-he^{-ix}}{1-he^{ix}}.
\ea
Note that, in this model,
the solutions of the Bethe ansatz equations
are restricted as $\mbox{Re}(k_j), \mbox{Re}(\Lam_\gamma)\geq 0$
and $k_j, \Lam_\gamma\ne 0$.
The energy of the model is represented as
\be
\label{energy}
E_N = -2 \sum_{j=1}^{N} \cos k_{j}.
\ee


Next, we shall briefly review the work of Grisaru et al. \cite{GMN}.
In \cite{K79}, 
Korepin gave a general method for exactly extracting 
the bulk $S$ matrix from the Bethe ansatz equations.
Then, generalizing this method, 
Grisaru et al. proposed the method for determining
the boundary $S$ matrix from the Bethe ansatz equations,
and applied this method to the open Heisenberg chain
with boundary magnetic fields \cite{GMN}.
Also, using this method,
Essler et al. calculated the boundary $S$ matrices
for the open supersymmetric $t$-$J$ model 
with boundary magnetic fields
and those for the open supersymmetric $t$-$J$ model 
with an impurity \cite{Ess96}.

An essential ingredient of their method 
is the following quantization condition \cite{FS94}
for a system of two particles, which has
the internal degrees of freedom,
with factorized scattering on a line of length $\tilde{L}$;
\be
\label{q-condition}
e^{2ip(\th_{1})\tilde{L}} 
S_{12}(\th_{1}-\th_{2})K_{1}(\th_{1},h_{1})
S_{12}(\th_{1}+\th_{2})K_{1}(\th_{1},h_{L})
=1,
\ee
where $\th_{j}$ is the rapidity of particle $j=1,2$, 
and $p(\th)$ is defined by the {\it expression}
for the momentum of a particle 
on the corresponding periodic system.
Here $S_{12}(\theta_1-\theta_2)$ is the (bulk) $S$ matrix
for the scattering of particles 1 and 2, 
and $K_{1}(\theta_1,h)$
is the  boundary $S$ matrix of the scattering for 
particle 1 off a boundary with boundary field $h$.
Under appropriate conditions 
on $S_{12}(\theta_1-\theta_2)$ and $K_{1}(\theta_1,h)$,
the equation (\ref{q-condition}) is equivalent to 
the following scalar equation (after taking logarithm);
\ba
\label{log-q-cond}
2\tilde{L}p(\th_{1})
&+&
\mbox{(bulk two-body phase shifts)} \nn \\
&+&
\mbox{(boundary phase shifts for $h_1$ and $h_L$)}
\equiv 0\ \ (\mbox{mod}\ \ 2\pi).
\ea
Note that, due to the factor $S_{12}(\theta_1+\theta_2)$
in eq. (\ref{q-condition}),
the bulk part of phase shifts
contains the phase shifts for  
the scattering of the particle 1
and the mirror image of particle 2.

On the other hand, if the system is Bethe ansatz solvable,
it is possible to derive 
the another condition on $p(\th_{1})$ from {\it the counting function} 
that is defined by the Bethe ansatz equations.
Then, comparing these two conditions,
the boundary phase shifts can be evaluated 
(up to rapidity independent constant) \cite{GMN}.


We now turn to consider 
the boundary scatterings of the open Hubbard chain.
 Since for the open Hubbard 
chain, it is reasonable to consider
the length of the system 
to be  $L+1$,
then we put $\tilde{L} =L+1$ 
in the discussions of the scatterings.

In this letter, we only consider the case 
with the bipartite lattice and the half filled band,
{\it i.e.}, $L$ even and $N=L$.
In this case, the elementary excitations of 
the periodic Hubbard Hamiltonian transform
in the fundamental representations of $SO(4)$ \cite{EK94,AN}.
These elementary excitations are called
spinons which carry spin but no charge
and holons/antiholons which carry charge but no spin 
\cite{Woy,EK94,AN}.
The excitation spectrum can be determined by
the scattering of these elementary excitations.
In ref. \cite{EK94,AN}, 
the bulk $S$ matrix for the periodic Hubbard chain
has been determined by using Korepin's method.
This $S$ matrix has the block diagonal form
with respect to the scattering of  
the spin excitations on the spin excitations, 
the spin excitations on the charge excitations,
the charge excitations on the spin excitations,
and the charge excitations on the charge excitations.

For the open Hubbard chain, the bulk part of 
the Hamiltonian is also $SO(4)$ invariant.
Thus, the elementary excitations are still
spinons and holons/antiholons.
However, in our choice of the Hamiltonian,
the $\eta$-$SU(2)$ symmetry is broken down to $U(1)$.
Thus the total $\eta$-spin is not a good quantum number.
The boundary $S$ matrices 
$K_{\rm spin}(\Lam,h)$ and $K_{\rm charge}(k,h)$
for spin and charge excitations, respectively,
have the following diagonal form,
since the Hamiltonian $H^{(+)}$ 
has $U(1)\times U(1)$ symmetry
which corresponds to the preservation of 
spinon and holon/antiholon numbers;
\ba
\label{bound-matrix-s}
K_{\rm spin}(\Lam,h) 
& = & 
\left(
 \begin{array}{cc}
  \cA(\Lam,h) & 0              \\
  0         & \cB(\Lam,h)
 \end{array}  
\right),        \\
\label{bound-matrix-c}
K_{\rm charge}(k,h) 
& = & 
\left(
 \begin{array}{cc}
  \cC(k,h) & 0          \\
  0      & \cD(k,h)
 \end{array}  
\right).       
\ea
Since the boundary Hamiltonian (\ref{boundary-hamil})
does not break the spin-$SU(2)$ symmetry,
we expect that the boundary $S$ matrix for the spin excitations 
is proportional to the identity matrix,
{\it i.e.}, $\cA(\Lam)=\cB(\Lam)$.
In fact, we will confirm this fact. 
Also we define the corresponding boundary phase shifts
by the formulae;
$\cA(\Lam,h)=e^{ia(\Lam,h)}$,
$\cB(\Lam,h)=e^{ib(\Lam,h)}$,
$\cC(k,h)=e^{ic(k,h)}$,
and
$\cD(k,h)=e^{id(k,h)}$.
{}From the same argument 
as was given by Grisaru et al. \cite{GMN},
to determine the above four components,
it is sufficient to analyze the highest weight states  
and the lowest weight states 
of the spin (resp. charge) excitation
with $S=1$ (resp. $\eta =1$).  
Here $S$ (resp. $\eta$) denotes 
the total spin (resp. $\eta$-spin) quantum number.
Note that hereafter we call 
the states which become $\eta=1$ states 
when the boundary fields vanish,
$\eta =1 $ states.
Notice also that, to study the scattering,
we can restrict attention to the states
near the ground state, {\it i.e.},
the states which have the microscopic
number of holes in the {\it real} roots. 

Let us introduce counting functions 
for roots $\{k_j\}$ and $\{\Lam_\g\}$ \cite{S85,AS,DY}.
As mentioned above,
for later purpose, we only need the real solutions
of the Bethe ansatz equations 
(\ref{bethe-eq1}) and (\ref{bethe-eq2}).
In this case, taking the logarithm of 
eq. (\ref{bethe-eq1}) and (\ref{bethe-eq2}),
we have
\ba
\label{log-bethe-eq1}
n_{j} 
& = & 
z_{c}(k_{j}), \\
\label{log-bethe-eq2}
I_{\gamma} 
& = & z_{s}(\Lam_{\g}), 
\ea
where $z_c(k)$ and $z_s(\Lam)$ are counting functions
for roots $\{k_j\}$ and $\{\Lam_\g\}$, respectively;
\ba
\label{count-c}
z_{c}(k) 
& = & 
\frac{1}{2\pi}
\Big[
2k \tilde{L}
+\frac{1}{i}\ln \beta(k,h_{1})
+\frac{1}{i}\ln \beta(k,h_{L}) \nn \\
&&
-\sum_{\dl=-M}^{M} \Theta (2\sin k-2\Lam_{\dl}) 
+\Theta(2\sin k)      
\Big],\\
\label{count-s}
z_{s} (\Lam) 
& = & 
\frac{1}{2\pi} 
\Big[
-\sum_{j=-N}^{N} \Theta (2 \Lambda -2 \sin k_{j} ) 
+\sum_{\delta = -M }^{M} \Theta (\Lambda - \Lambda_{\delta}) 
-\Theta(\Lambda)
\Big],
\ea
with $ \Theta (x)  =  -2 \tan^{-1} ( x/c) $.
In the above expressions, we have used the 'doubling trick',
that is, we have put $\Lambda_{-\delta} = -\Lambda_{\delta},  
k_{-j} = -k_{j}$, and $\Lam_0, k_0=0$. 
The two sequences of quantum numbers 
$\{n_{j}\}_{j=1}^N$ and $\{I_{\g}\}_{\g=1}^M$ 
(we call $n$-sequence and $I$-sequence respectively)
take values in integers,
and label the state of the model.
Remark that $n_{j}$'s, which are defined modulo $2\tilde{L}$,
take values in $0<n_{j}\leq N$.
Also remark that, from the formula $|\Theta(x)|\leq \pi$, 
$I_{\gamma}$'s are restricted as 
$0<I_{\gamma}\leq N-M(=I_{\mbox{{\scriptsize max}}})$.
For instance,
the ground state is characterized by $M = N/2$ (spin singlet)
and the configuration $n_{j}=j, I_{\gamma}=\gamma$.


We shall also introduce the densities of roots and holes.
The number of allowed solutions for the Bethe ansatz equations 
(\ref{bethe-eq1}) and (\ref{bethe-eq2}) 
in the intervals ($k$, $k+dk$) and ($\Lam$, $\Lam+d\Lam$)
are expressed as 
$\tilde{L}[\rho(k)+\rho^{h}(k)]dk$ and 
$\tilde{L}[\sigma(\Lam)+\sigma^{h}(\Lam)]d\Lam$.
Here $\rho(k)$ and $\sigma(\Lam )$ 
are the densities of roots (filled solutions),
and $\rho^{h}(k)$ and $\sigma^{h}(\Lam)$ 
are the densities of holes (unfilled solutions).
These are determined by the counting functions as follows;
\ba
\label{den-c}
\tilde{L}[\rho (k)+\rho^{h}(k)] 
& =  & d z_{c} (k)/dk,  \\
\label{den-s}
\tilde{L}[\sigma(\Lam )+\sigma^{h}(\Lam)] 
& = & d z_{s} (\Lam )/d\Lam.
\ea
Conversely, we can determine the counting functions from 
the integration of the above formulae,
if we know the explicit form of 
$\rho (k), \rho^{h}(k), \sigma(\Lam )$, and $\sigma^{h}(\Lam)$.

In the thermodynamic limit 
($\tilde{L} \rightarrow \infty$ with $ N/\tilde{L}$ and $M/\tilde{L}$ fixed), 
we obtain the following formulae;
\ba
\label{den-count-c}
\rho(k)+\rho^{h}(k) 
& = & 
\frac{1}{\pi} 
+2\cos k\int_{-B}^{B} d\Lam
\sigma(\Lam)K(2\sin k-2\Lam) \nn \\
&&
+\frac{1}{\tilde{L}\pi}[\tau(k,h_{1})+\tau(k,h_{L})]
-\frac{2\cos k}{\tilde{L}}K(2\sin k), \\
\label{den-count-s}
\sigma(\Lam)+\sigma^{h}(\Lam) 
& = & 
2\int_{-Q}^{Q} dk\rho(k)K(2\Lam-2 \sin k)
-\int_{-B}^{B} d\Lam' \sigma(\Lam') K (\Lam-\Lam')\nn \\
&&
+\frac{1}{\tilde{L}}K(\Lam), 
\ea
where 
$K(x) = c/[\pi (x^{2} +c^{2})]$ 
and
$\tau(x,h)=(h\cos x-h^{2})/(1-2h\cos x+h^2)$.
Here the charge and spin  pseudo Fermi-momenta $Q$ and $B$, 
respectively, are determined by the conditions 
\ba
\label{def-Q}
\int_{-Q}^{Q} dk \rho(k) & = & (2N+1)/\tilde{L}, \\
\label{def-B}
\int_{-B}^{B} d\Lam \sigma(\Lam) & = & (2M+1)/\tilde{L}.
\ea

Since we have to determine the densities of order $1/\tilde{L}$, 
we may expand $\rho(k)$ and $\sigma(\Lam)$ as
\ba
\label{1/L-exp-c}
\rho(k) 
& = & 
\rho_{0}(k)+\rho_{1}(k)/\tilde{L}
+O(1/\tilde{L}^2),\\
\label{1/L-exp-s}
\sigma(\Lam ) 
& = & 
\sigma_{0}(\Lam)+\sigma_{1}(\Lam)/\tilde{L}
+O(1/\tilde{L}^2).
\ea 
For example, we can easily derive the ground state densities 
$\rho_0^{(g)} (k)$ and $\sigma_0^{(g)}(\Lam)$ 
of order $ O(\tilde{L}^{0})$.
In the ground state there are no holes.
Also, in the half filling case, we see 
$Q=\pi$ and $B= \infty$ for order $ O(\tilde{L}^{0})$.
Then we can solve 
eqs. (\ref{den-count-c}) and (\ref{den-count-s});
\ba
\label{ground-den-c}
\rho_0^{(g)} (k) & = & \frac{1}{\pi} +\frac{\cos k}{2\pi} 
\int_{-\infty}^{\infty} dp 
\frac{J_{0} (p) e^{-ip\sin k -c|p|/2} }{\cosh (cp/2)},\\
\label{ground-den-s}
\sigma_0^{(g)} (\Lam) & = & \frac{1}{2\pi} 
\int_{-\infty}^{\infty} dp \frac{J_{0}(p)e^{-ip\Lam}}{\cosh (cp/2)},
\ea
where $J_{0}(p)$ is the zeroth-order Bessel function.

To determine the boundary $S$ matrices for the open Hubbard chain,
we now proceed to study the excitations 
which are classified by $S_{z}$ and $\eta_{z}$ eigenvalues.
Here $S_{z}$ (resp. $\eta_{z}$) denotes 
the total $z$-component of spins (resp. $\eta$-spins).
We will consider the spin and charge excitations separately.

\noindent
{\it Spin excitation}

We consider the spin excitations. 

We first investigate the state with $S=1, S_{z} = 1$.
{}From this excitation, we can determine 
the component $\cA(\Lam,h)$ 
of the boundary $S$ matrix $K_{\rm spin}(\Lam,h)$.

The $S=1, S_{z} =1$ state is obtained 
by $M \rightarrow N/2 -1$ with $N$ fixed.
In this case 
$I_{\mbox{{\scriptsize max}}} =N-M = N/2 +1$.
Thus there are two holes $I_{1}^{h},I_{2}^{h}$ in the $I$-sequence,
and the $n$-sequence do not change. 
We denote the corresponding spin rapidities 
$\Lam_{\alpha}^{h}$ ($\alpha =1,2$), that is, 
$ I_{\alpha}^{h} = z_{s}(\Lam_{\alpha}^{h})$.
The hole densities are thus given by
\ba
\label{s-ex-hole-den-c}
\rho^{h}(k)
&=&0,\\
\label{s-ex-hole-den-s}
\sigma^{h}(\Lam) 
& = & 
\frac{1}{\tilde{L}}
[ 
 \delta(\Lam-\Lam_{1}^{h})
+\delta(\Lam+\Lam_{1}^{h})
+\delta(\Lam-\Lam_{2}^{h})
+\delta(\Lam+\Lam_{2}^{h}) 
].
\ea
Then we obtain the integral equations for pairs
$(\rho_{0} (k), \sigma_{0} (\Lam))$ and 
$(\rho_{1} (k), \sigma_{1} (\Lam))$
with integration boundaries $Q$ and $B$ 
which are defined by eqs. (\ref{def-Q}), (\ref{def-B}). 
Since we determine the densities of order $O(1/\tilde{L})$,
the shifts of the integration boundaries from 
the ground state must be examined of order $O(1/\tilde{L})$.
Following refs. \cite{FS94,GMN,Ess96},
we assume that, in the thermodynamic limit, 
the shifts of the integration boundaries 
are of order $O(1/\tilde{L}^n), (n\geq 2)$, 
as far as the boundary phase shifts are concerned.
Under this assumption, integral equations can be solved.
We then obtain $\rho_{0} (k)=\rho_0^{(g)}(k), 
\sigma_{0} (\Lam)=\sigma_0^{(g)}(\Lam)$, and 
\ba
\rho_{1}(k) 
& = & 
-\frac{\cos k}{2c} 
\left\{ 
  \frac{1}{\cosh [\pi (\sin k -\Lam_{1}^{h})/c]}
+ \frac{1}{\cosh [\pi (\sin k +\Lam_{1}^{h})/c]}
\right.  \nn \\
&  & \ \ \ \ \ \ \ \ \ 
\left.
+ \frac{1}{\cosh [\pi (\sin k -\Lam_{2}^{h})/c]}
+ \frac{1}{\cosh [\pi (\sin k +\Lam_{2}^{h})/c]} 
\right\} \nn \\
&   & +
\frac{\cos k }{4\pi^{2}} 
\int_{-\infty}^{\infty} dp
\int_{-\pi}^{\pi} dk' 
[\tau(k',h_{1}) +\tau(k',h_{L})]
\frac{e^{-2ip (\sin k -\sin k')-c |p| }}{\cosh (cp)}\nn \\
&   & + 
\frac{\cos k}{2\pi} \int_{-\infty}^{\infty} dp 
\frac{e^{-ip \sin k -c|p| /2}}{1+e^{c|p|}} \nn \\
&   & +
\frac{1}{2\pi}[\tau(k,h_{1}) +\tau(k,h_{L})]
-2\cos k K (2\sin k), \\
\sigma_{1} (\Lam) 
& = & 
-\frac{1}{\pi} \int_{-\infty}^{\infty} dp 
[\cos (p\Lam_{1}^{h})+\cos(p \Lam_{2}^{h})]
\frac{e^{-ip\Lam}}{1+e^{-c|p|}}\nn \\
&   & +
\frac{1}{4\pi c} \int_{-\pi}^{\pi} dk 
 \frac{\tau(k,h_{1}) +\tau(k,h_{L})}
{\cosh [\pi (\sin k -\Lam )/c ] } \nn \\
&   & +
\frac{1}{2\pi} 
\int_{-\infty}^{\infty} dp 
\frac{e^{-ip\Lam -c|p|}}{1+e^{-c|p|}}.
\ea

{}From the above formulae,
we can obtain the following equation for 
the counting function $z_{s}(\Lam)$ in the thermodynamic limit;
\ba
\label{count-s-at-hole}
-2\pi z_{s}(\Lam_{1}^{h}) 
& = & 
2 \tilde{L}p_{s} (\Lam_{1}^{h} )
+{\cal N}_1(\Lam_{1}^{h})
+{\cal N}_2(\Lam_{1}^{h},\Lam_{2}^{h})
\equiv 0 \ \ (\mbox{mod} \ \ 2\pi),
\ea
where $p_{s} (\Lam_{1}^{h})$ is defined by the expression 
for the spinon momentum of the corresponding periodic system
\cite{Woy,EK94,AN};
\be
p_{s} (\Lam_{1}^{h})
=
-\int_0^\infty
\frac{dp}{p}\frac{J_0(p)}{\cosh(cp/2)}\sin(p\Lam_{1}^{h}).
\ee
Terms ${\cal N}_{1}(\Lam_{1}^{h})$ 
and ${\cal N}_2(\Lam_{1}^{h},\Lam_{2}^{h})$ 
in (\ref{count-s-at-hole}) are given by 
\ba
\label{bound-phase-s}
{\cal N}_1(\Lam_{1}^{h})
&=&
 \gamma(-2\Lam_{1}^{h}/c)
+\gamma(-\Lam_{1}^{h}/c) \nn \\
&& 
-\frac{1}{2\pi} 
\int_{-\pi}^{\pi} dk
[\tau(k,h_{1}) +\tau(k,h_{L})] \nn \\
&& \ \ \ \ \ \ \ \ \ \ \ \ \ \ 
\times\phi(-i(\Lam_{1}^{h} -\sin k)/c,
1/4+i(\Lam_{1}^{h} -\sin k)/2c), \\
\label{bulk-phase-s}
{\cal N}_2(\Lam_{1}^{h},\Lam_{2}^{h})
&=&
 \gamma(-(\Lam_{1}^{h}-\Lam_{2}^{h})/c)
+\gamma(-(\Lam_{1}^{h}+\Lam_{2}^{h})/c),
\ea
where
\ba
\phi(x,y) 
& = & 
i\int_{0}^{\infty} \frac{d\omega}{\omega}
\frac{(1-e^{-2x \omega})e^{-2y \omega}}{1+e^{-\omega}} \nn \\
 & = & 
i\ln\frac{\Gamma(x+y+1/2)\Gamma(y)}
         {\Gamma(x+y)\Gamma(y+1/2)},  \\
\gamma(x) & = & -\phi (ix , (1-ix)/2) \nn \\
 & =& 
i\ln\frac{\Gamma((1-ix)/2)\Gamma(1+ix/2)}
         {\Gamma((1+ix)/2)\Gamma(1-ix/2)}. 
\ea
We see that 
${\cal N}_2(\Lam_{1}^{h},\Lam_{2}^{h})$
are the bulk phase shifts due to the scatterings of 
the particle 1 and 2 , 
and also the particle 1 and the mirror image of the particle 2.
Similarly, we can conclude  
that ${\cal N}_{1}(\Lam_{1}^{h})$ 
is the sum of boundary phase shifts 
for the scattering of particle 1 off boundaries 
with boundary fields $h_1$ and $h_2$.
That is, ${\cal N}_{1}(\Lam)=a(\Lam,h_1)+a(\Lam,h_2)$.
Therefore we determine $\cA(\Lam,h)$ 
up to the rapidity independent constant.


To calculate the component ${\cal B}(\Lam,h)$ in the 
equation (\ref{bound-matrix-s}), 
we next consider the state with $S=1, S_{z} =-1$.
We find that, for this spin-$SU(2)$ invariant case,
the Bethe ansatz equations and energy spectrum 
of the $S=1, S_{z} =-1$ state 
are trivially same as those for the $S=1, S_{z} =1$ state.
Thus we have ${\cal A}(\Lam,h)={\cal B}(\Lam,h)$.

\noindent
{\it Charge excitation}

Next we consider the  charge excitations. 
To determine the two component of the boundary $S$ matrix 
for the charge excitations, 
we must consider the $\eta =1, \eta_{z} =1$ state and the 
$\eta =1, \eta_{z} =-1$ state by the Bethe ansatz.

The $\eta =1, \eta_{z} =-1$ state is obtained 
by removing two $k$'s from the ground state,
{\it i.e.}, $N=L-2$ and $M=N/2$.
In this case, we have
\ba 
\sigma^{h}(\Lam) 
& = & 0,\\
\rho^{h} (k)  
& = & \frac{1}{\tilde{L}} 
[ 
  \delta(k-k_{1}^{h}) 
+ \delta(k+k_{1}^{h}) 
+ \delta(k-k_{2}^{h}) 
+ \delta(k+k_{2}^{h})
].
\ea
Similar to the case of the spin excitation,
under the assumption for the integration boundaries,
we obtain the densities 
$\sigma_{0}(\Lam), \rho_{0}(k), \sigma_{1}(\Lam)$ and $\rho_{1}(k)$.
Results are $\rho_{0} (k)=\rho_0^{(g)}(k), 
\sigma_{0} (\Lam)=\sigma_0^{(g)}(\Lam)$, and
\ba
\rho_{1}(k) 
& = & 
-\tilde{L}\rho^{h} (k) 
-\frac{\cos k}{2\pi} \int_{-\infty}^{\infty} dp 
[\cos(p\sin k_{1}^{h})+\cos(p\sin k_{2}^{h}) ]
\frac{e^{-c|p|/2 }}{\cosh (cp/2)} \nn \\
&   & 
+\frac{\cos k}{8\pi^2} 
\int_{-\infty}^{\infty} dp
\int_{-\pi}^{\pi} dk' 
[\tau(k',h_{1}) +\tau(k',h_{L})] 
\frac{e^{-ip(\sin k -\sin k')-c|p|}}{\cosh(cp/2)} \nn \\
& &
+\frac{\cos k}{2\pi} \int_{-\infty}^{\infty} dp
\frac{e^{-ip\sin k-c|p|/2}}{1+e^{c|p|}}\nn \\
& & 
+\frac{1}{2\pi}[\tau(k,h_{1}) +\tau(k,h_{L})], \\
\sigma_{1} (\Lam) 
& = & 
-\frac{1}{2c} 
\left\{ 
 \frac{1}{\cosh[\pi(\Lam-\sin k_{1}^{h})/c]}
+\frac{1}{\cosh[\pi(\Lam+\sin k_{1}^{h})/c]}\right. \nn \\
 & & \ \ \ \ \ \ 
\left.
+\frac{1}{\cosh[\pi(\Lam-\sin k_{2}^{h})/c]}
+\frac{1}{\cosh[\pi(\Lam+\sin k_{2}^{h})/c]} 
\right\} \nn \\
 & & +\frac{1}{4\pi c} \int_{-\pi}^{\pi} dk
\frac{\tau(k,h_{1}) +\tau(k,h_{L})}
     {\cosh[\pi(\sin k-\Lam)/c]} \nn \\
& &
+ \frac{1}{2\pi} \int_{-\infty}^{\infty} dp 
\frac{e^{-ip\Lam}}{1+e^{c|p|}} .
\ea
Also, we have the counting function in the thermodynamic limit
\ba
-2\pi z_{c} (k_{1}^{h}) 
=
2\tilde{L}p_{c}^{\eta_{z}=-1}(k_{1}^{h})
+{\cal M}_{1} (k_{1}^{h} )
+{\cal M}_{2} (k_{1}^{h},k_{2}^{h}) 
\equiv 0 \ \ (\mbox{mod} \ \ 2\pi),
\ea
where $p_{c}^{\eta_{z} =-1}(k_{1}^{h})$
is the quasiparticle momentum of the corresponding periodic system
\cite{Woy,EK94,AN};
\ba
p_{c}^{\eta_{z} =-1}(k_{1}^{h}) 
& = & 
-k_{1}^{h} 
-\int_{0}^{\infty} \frac{dp}{p} 
\frac{J_{0}(p)e^{-cp/2}}{\cosh (cp/2)} 
\sin (p\sin k_{1}^{h}),
\ea
and 
\ba 
{\cal M}_{1} (k_{1}^{h} ) 
& = & 
\gamma (-2\sin k_{1}^{h}/c)
-2k_{1}^{h} -\frac{1}{i} 
[\ln \beta(k_{1}^{h},h_{1}) 
+\ln \beta(k_{1}^{h},h_{L})]  \nn \\
 & & 
-\phi(i\sin k_{1}^{h}/c,3/4-i\sin k_{1}^{h}/c) 
-\Theta (2\sin k_{1}^{h}) \nn \\
& & -\frac{1}{2\pi}\int_{-\pi}^{\pi} dk'
[\tau(k',h_{1}) +\tau(k',h_{L})]
\gamma (-(\sin k'-\sin k_{1}^{h})/c), \\
{\cal M}_{2} (k_{1}^{h},k_{2}^{h}) 
& = & 
 \gamma(-(\sin k_{1}^{h}-\sin k_{2}^{h})/c)
+\gamma(-(\sin k_{1}^{h}+\sin k_{2}^{h})/c).
\ea
Then, we can determine the component $\cD(k,h)$ from  
${\cal M}_{1} (k_{1}^{h})$.


Finally, we have to determine the remaining 
component ${\cal C}(k,h)$ in (\ref{bound-matrix-c}).
Let us study the $\eta=1, \eta_{z} =1$ state 
to determine $\cC(k,h)$. 
Since the  $\eta=1, \eta_{z} =1$ state is not the regular Bethe
ansatz state \cite{EKS92}, 
we must take the completely filled
state $|\Omega\rangle =\prod_{j=1}^{L} \psi_{j \uparrow}^{\dagger} 
\psi_{j \downarrow}^{\dagger} |0\rangle$ as the
Bethe ansatz vacuum. 
The Bethe ansatz state with $2L-N$ electrons 
is thus given as 
\ba
| \Phi_{N}\rangle 
& = & 
\sum_{\sigma_{1},\cdots,\sigma_{N}\in\{\uparrow,\downarrow\}}
 \Phi_{\sigma_{1},\cdots,\sigma_{N}} (n_{1},\cdots,n_{N}) 
\prod_{i=1}^{N} \psi_{n_{i}\sigma_{i} } |\Omega \rangle,
\ea 
where $n_{i}$'s denote the location of electrons on the chain.
It is easy to see that
the eigenvalue of the Hamiltonian  $H^{(+)}$ for this state 
is given by $E_N'=-E_N$,
and the Bethe ansatz equations are obtained by taking 
$c\rightarrow-c$
in the equations (\ref{bethe-eq1}) and (\ref{bethe-eq2}).
Then the problem reduces to find the eigenstates of 
the attractive Hubbard model with the eigenvalues 
which are given by changing those signs from the 
corresponding eigenvalues for the repulsive case.
That is, the ground state configuration of rapidities
for our model
is identical to the highest energy configuration of rapidities
for the attractive case. 
This is the configuration that all rapidities are real 
and $N=L, M=N/2$ \cite{W832}. 
Therefore, the $\eta =1,\eta_{z}=1$ state is obtained 
by removing two $k$'s from the ground state configuration.
Repeating the calculation similar to the case of 
the $\eta=1, \eta_{z} =-1$ state, we have 
\ba
2\pi z_{c} (k_{1}^{h}) 
=
2\tilde{L}p_{c}^{\eta_{z} =1}(k_{1}^{h})
+{\cal M}_{1}'(k_{1}^{h})
+{\cal M}_{2}'(k_{1}^{h},k_{2}^{h})
\equiv 0 \ \ (\mbox{mod} \ \ 2\pi), 
\ea
where $p_{c}^{\eta_{z} =1}(k_{1}^{h})$ 
is the quasiparticle momentum
(note that  
$p_{c}^{\eta_{z} =1} $ is different to 
$p_{c}^{\eta_{z} =-1} $);
\ba
p_{c}^{\eta_{z} =1}(k_{1}^{h}) & = & k_{1}^{h} 
-\int_{0}^{\infty} \frac{dp}{p} 
\frac{J_{0} (p) 
e^{-cp/2} }{\cosh (cp/2) } 
\sin (p\sin k_{1}^{h} ),
\ea
and 
\ba 
{\cal M}_{1}'(k_{1}^{h}) & = & 
\gamma (-2\sin k_{1}^{h}/c)
+ 2k_{1}^{h} + \frac{1}{i} 
[\ln \beta(k_{1}^{h},h_{1}) 
+\ln \beta(k_{1}^{h},h_{L})]  \nn \\
&&
-\phi (i\sin k_{1}^{h}/c,3/4-i\sin k_{1}^{h}/2c) 
-\Theta (2\sin k_{1}^{h}) \nn \\
& & -\frac{1}{2\pi}\int_{-\pi}^{\pi} dk'
[\tau(k',h_{1}) +\tau(k',h_{L})]
\gamma (-(\sin k'-\sin k_{1}^{h})/c), \\
{\cal M}_{2}'(k_{1}^{h},k_{2}^{h}) 
& = & 
 \gamma (-(\sin k_{1}^{h}-\sin k_{2}^{h})/c)
+\gamma (-(\sin k_{1}^{h}+\sin k_{2}^{h})/c).
\ea

As for the case of the $\eta=1, \eta_{z} =-1$ state,
we obtain  the component $\cC(k,h)$ from 
${\cal M}_{1} (k_{1}^{h})$.

\noindent
{\it Boundary $S$ matrices}

Now let us summarize the results.
Up to rapidity-independent phase factors,
the resulting boundary $S$ matrices are expressed as
\ba
K_{\rm spin}(\Lam,h) 
& =&  
e^{{\cal X}_s(\Lam,h)}
\left( 
 \begin{array}{cc}
   1 & 0 \\
   0 & 1
 \end{array} 
\right),\\
K_{\rm charge}(k,h) 
& =&  
e^{{\cal X}_c(k,h)}
\left( 
 \begin{array}{cc}
    \beta(k,h) & 0                     \\
   0                 &  \beta(k,h)^{-1} 
 \end{array} 
\right),
\ea
where 
\ba
\label{phase-s}
2{\cal X}_s(\Lam,h)
&=& 
\gamma(-2\Lam/c)+\gamma(-\Lam/c)  \nn \\
& & 
-\frac{1}{\pi} \int_{-\pi}^{\pi} dk\tau(k,h)
\phi(-i(\Lam-\sin k)/c,1/4+i(\Lam-\sin k)/2c),\\
\label{phase-c}
2{\cal X}_c(k,h) 
& = & 
\gamma (-2\sin k/c)\nn \\
& &  
- \phi (i\sin k/c,3/4-i\sin k/2c) 
- \Theta (2\sin k) \nn \\
&&
-\frac{1}{\pi}\int_{-\pi}^{\pi} dk'\tau(k',h)
\gamma (-(\sin k'-\sin k)/c).
\ea

It is noteworthy that,
in contrast to the case of 
open supersymmetric $t$-$J$ model \cite{Ess96}, 
the boundary $S$ matrix for the spin excitations
depends on the boundary field 
although the boundary field does not break the spin-$SU(2)$ symmetry.

If the boundary fields vanish, the boundary $S$ matix of the
charge excitations becomes proportional to the identity matrix
as expected.
The bulk $S$ matrices for the Hubbard chain \cite{EK94}\cite{AN} 
and the supersymmetric $t$-$J$ model \cite{Ess96} have the
same form as that for the XXX chain.
However the boundary $S$ matrix for the open Hubbard chain
has different form with the one for the open XXX model.
Full details and applications of our results
will be published elsewhere.
 
\vspace{12pt}
 
\noindent {\bf Acknowledgment:} 
I wish to thank Dr. T. Yamamoto for discussions and comments.
I am also indebted to Prof. A. Kuniba for  discussions. 
 
\vspace{24pt}


\end{document}